\documentclass[aps,prl,twocolumn,amsmath,superscriptaddress,floatfix]{revtex4-1}

\usepackage[version=3]{mhchem}
\usepackage{graphicx}
\usepackage{dcolumn}
\usepackage{bm}
\usepackage{epstopdf}
\usepackage{amssymb}
\usepackage{xcolor}
\usepackage{nicefrac}
\usepackage{bbm}

\usepackage{float}

\begin{document}
\title{Sensing individual nuclear spins with a single rare-earth electron spin}

\author{Thomas Kornher}
\email{t.kornher@pi3.uni-stuttgart.de}
\affiliation{3rd Institute of Physics, University of Stuttgart, 70569 Stuttgart, Germany}

\author{Da-Wu Xiao}
\affiliation{Beijing Computational Science Research Center, Haidian District, Beijing 100194, China}

\author{Kangwei Xia}
\affiliation{3rd Institute of Physics, University of Stuttgart, 70569 Stuttgart, Germany}

\author{Fiammetta Sardi}
\affiliation{3rd Institute of Physics, University of Stuttgart, 70569 Stuttgart, Germany}

\author{Nan Zhao}
\affiliation{Beijing Computational Science Research Center, Haidian District, Beijing 100194, China}
\author{Roman Kolesov}
\affiliation{3rd Institute of Physics, University of Stuttgart, 70569 Stuttgart, Germany}

\author{J\"org Wrachtrup}
\affiliation{3rd Institute of Physics, University of Stuttgart, 70569 Stuttgart, Germany}

\title{Sensing individual nuclear spins with a single rare-earth electron spin}


\begin{abstract} 
Rare-earth
related electron spins in crystalline hosts are
 unique material systems, as they can potentially provide a direct interface between telecom band 
 photons and long-lived spin quantum bits.
Specifically, their optically accessible electron spins in solids interacting with nuclear spins in 
their environment
are valuable quantum memory resources. Detection of nearby individual nuclear spins,
so far exclusively shown for few dilute nuclear spin bath host systems such as the
NV center in diamond or the silicon vacancy in silicon carbide, remained an 
open challenge for rare-earths in their host materials, which typically exhibit dense nuclear spin baths. 
Here, we present
the electron spin
spectroscopy of single Ce$^{3+}$ ions in a yttrium orthosilicate host, featuring a
coherence time of $T_{2}=124\,\mu$s. This coherent interaction time is sufficiently long
to isolate proximal $^{89}$Y nuclear spins
from the nuclear spin bath of $^{89}$Y. 
Furthermore, it allows for the
detection of a single nearby $^{29}$Si nuclear spin, native to the host material with 
\raisebox{-0.9ex}{\~{}}5\,\% abundance. This study opens the door to quantum memory
applications in rare-earth ion related systems based on coupled environmental nuclear spins, potentially 
useful for quantum error correction schemes. 
\end{abstract}
\maketitle

Hybrid quantum systems, consisting of a read-out electron spin and long-lived
nuclear spins, have demonstrated remarkable properties for quantum memory applications
 \cite{morton2008solid,maurer2012room, awschalom2018quantum}. 
 At the same time single spins enable active quantum processing used in e.g. 
 quantum error correction 
\cite{waldherr2014quantum}. Implementing them in scalable quantum networks based on single
rare-earth ions (REI)
doped in solids potentially combines long distance entanglement distribution via single telecom band 
photons\cite{dibos2018atomic} with error corrected long-lived quantum memories. 
 \begin{figure}[htp]
 \begin{center}
\includegraphics[width=\columnwidth]{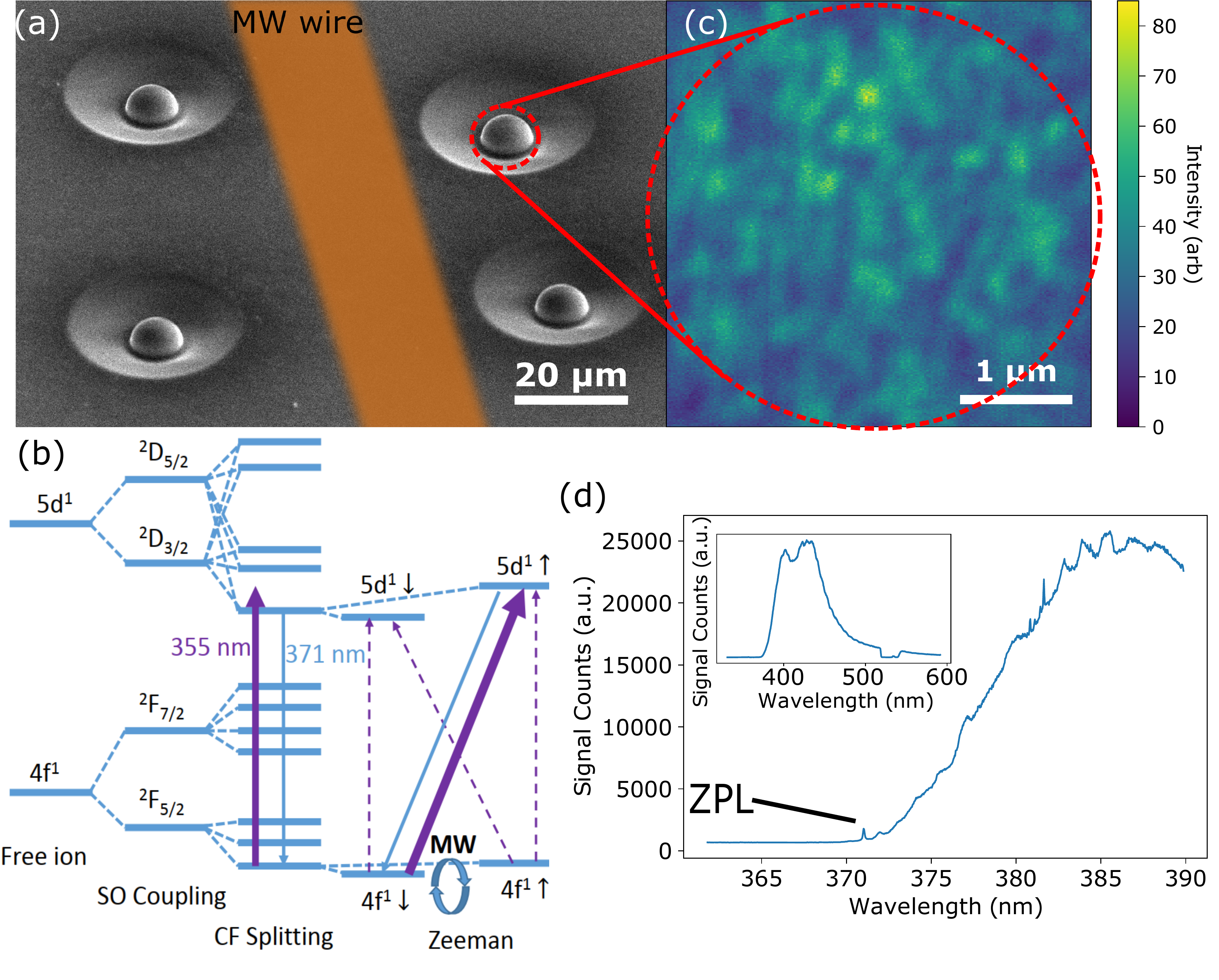}
\caption{(a) SEM image of SILs on a YSO crystal surface.
The MW copper wire position is indicated by an orange stripe.
(b) Electronic level structure of Ce$^{3+}$ ion in YSO crystal. Purple arrows indicate 
laser excitation and blue arrows fluorescence. Under circularly polarized excitation light, directed along
the same axis as the applied external magnetic field (along the b axis of the crystal), different
strengths of optical transitions between the lowest $4f$ level and the lowest $5d$ level are indicated by 
different widths of arrows. (c) Confocal scan of cerium centers under pulsed laser excitation, 
taken at the focus of a SIL.
(d) High-resolution spectrum of 0.01\,\% Ce$^{3+}$:YSO crystal in the vicinity of the 
zero-phonon line at 371\,nm. Inset: Typical spectrum of a 0.01\,\% Ce$^{3+}$:YSO crystal at 8\,K.
The gap in the spectrum around 532\,nm is an artifact of the used notch filter.
\label{fig:img1}}
\end{center}
\end{figure}

Based on the efficient isolation of REI's $4f$ electrons, their 
narrow and stable optical and spin levels have been used for
demonstration of storage and retrieval of single photons
\cite{de2015quantum} and exceptional coherence times \cite{zhong2015optically}, rendering them
particularly suitable for quantum repeater protocols.
With additional control over single rare-earth ions, 
however, these capabilities can be extended to high-fidelity spin readout and 
generation of entanglement of rare-earth electrons and nuclei in a scalable fashion\cite{awschalom2018quantum}.
Consequently, an increasing number of REI are isolated as single emitters 
 \cite{kolesov2012optical,yin2013optical,kolesov2013mapping,Zhong2018optically,dibos2018atomic}. 
 Based on ancillary electron spins of these single emitters, sensing of nuclear spins 
 \cite{zhao2012sensing,Kolkowitz2012sensing,Taminiau2012detection, car2018selective, zhong2019quantum} 
 is an important next
 step for REI based quantum network applications. 
So far, only dilute nuclear spin bath host materials such as 
 diamond \cite{childress2006coherent}, silicon carbide \cite{nagy2019high} and silicon \cite{pla2013high}
 were successfully used for 
 detection of individual nuclear spins.
 
In this study, we use the yttrium orthosilicate (Y$_2$SiO$_5$,YSO) crystal to
investigate the nulcear environment of individual Ce$^{3+}$ electron spins, which simultaneously act as a proxy
 for other REI species, owing to their interchangable doping into 
 yttrium containing solids.
We demonstrate spin initialization and coherent manipulation of Ce$^{3+}$ electron spins
 in a 
YSO host crystal.
Surprisingly we were able to sense individual dipolar coupled $^{29}$Si nuclear spins
despite the strong yttrium spin bath. $^{29}$Si signal is distilled 
 furthermore with basic decoupling sequences.
 Signatures of 
 yttrium nuclear spins also reveal dipolar coupling with the nearby Ce$^{3+}$
 superimposed on the yttrium spin bath.

Trivalent cerium substitutes Y$^{3+}$ in 95\,\% of the cases at the 7-oxygen-coordinated
site of the YSO crystal with a $C_1$ symmetry \cite{pidol2006epr}. 
The remaining 5\,\% of Ce$^{3+}$ ions substituting Y$^{3+}$ in the 6-oxygen-coordinated site 
of the crystal can be neclegted from further considerations 
due to different optical
(red shifted) and magnetic (different g-tensor)
properties \cite{pidol2006epr, drozdowski2004vuv}.
Individual Ce$^{3+}$ ions 
were identified in an ultra-pure YSO crystal using laser scanning confocal 
microscopy. The experimental setup of the microscope is described in 
detail in the supplementary material \cite{supplement}. 
To improve the collection efficiency and spatial resolution 
of the microscope, solid immersion lenses (SIL) were fabricated 
on the surface of the sample by focused ion beam milling. A scanning electron microscopy (SEM) image
of the sample with milled SILs and an indicated wire used for microwave (MW) spin
manipulation is shown in Fig. \ref{fig:img1}a. 
A picosecond pulsed laser at 355\,nm wavelength was used to off-resonantly excite 
Ce$^{3+}$ ions from the $4f$ ground state into the excited $5d$ band based on their phonon-assisted 
absorption band \cite{drozdowski2004vuv}. The corresponding energy level structure of Ce$^{3+}$ in YSO is
shown in Fig. \ref{fig:img1}b. For the purpose of acquiring a fluorescence spectrum, 
$5d \rightarrow 4f$ emission was collected from a 0.01\% doped Ce$^{3+}$:YSO crystal at $T\approx 8$\,K
(inset Fig. \ref{fig:img1}d).
 A high resolution spectrum of Ce$^{3+}$:YSO, shown in Fig. \ref{fig:img1}d, reveals the
zero-phonon line at 371\,nm, characteristic for Ce$^{3+}$ ion fluorescence at cryogenic temperatures 
\cite{yan2013measurement}. Performing confocal
microscopy with the pulsed 355\,nm laser on the SILs, however, allows 
for resolving individual Ce$^{3+}$ ions. A typical
laser scanning fluorescence image of optically resolved Ce$^{3+}$ ions located in the focus of a SIL is
shown in Fig. \ref{fig:img1}c.

 Even in ultra-pure crystals (from Scientific Materials), cerium is an
unavoidable native
impurity for yttrium-based hosts and the estimated residual density of 0.3\,ppb in our crystal is the
main contribution to background signal and can result in more than one Ce$^{3+}$ ion to be probed within
the focal volume. 
In principle, stimulated emission depletion (STED) based 
superresolution microscopy is available for Ce$^{3+}$:YSO (see supplementary material \cite{supplement}) 
\cite{kolesov2018superresolution}, however the
spin initialization and readout was found to be prevented by the high power depletion laser used in
the experiment.

  \begin{figure*}[htb]
\includegraphics[width=\textwidth]{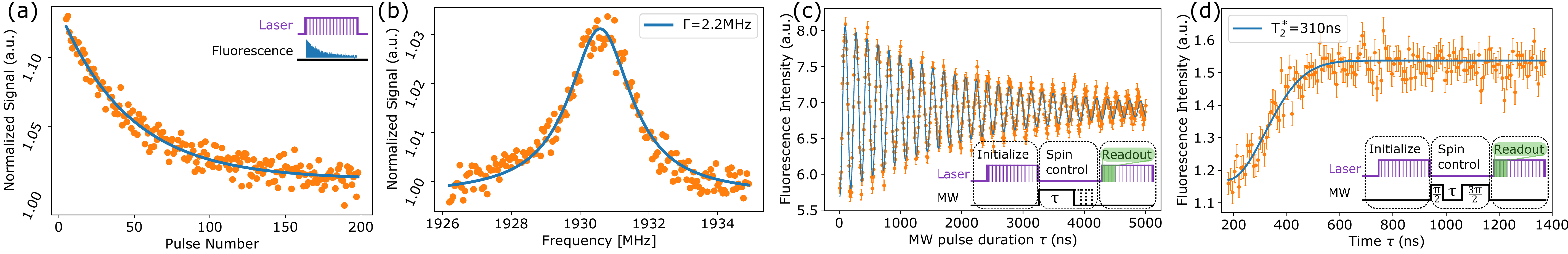}
\caption{a) Fluorescence intensity as function of the laser pulse number during
initialization. For coherent spin manipulation experiments, 100 pulses were used for spin initialization.
b) ODMR measurement of Ce$^{3+}$ as a function of MW frequency. A fitted
Lorentzian reveals a linewidth of 2.2\,MHz. 
c) Ce$^{3+}$ spin Rabi oscillations. Blue line is a fit to the data with an exponentially decaying 
cosine function with constant offset. d) Ramsey measurement of Ce$^{3+}$ 
with Gaussian fit yielding $T_{2}^{*}=310$\,ns.
\label{fig:img2}}
\end{figure*}
In a magnetic field parallel to the optical excitation beam, optical
transitions between $4f^1$ and $5d^1$ spin doublets show different relative strengths under 
$\sigma^+$ circularly polarized excitation \cite{kolesov2013mapping}. 
While the $|4f^1 \downarrow \rangle \rightarrow |5d^1 \uparrow \rangle$ spin-flip transition exhibits
the strongest optical dipole moment out of the four possible transitions 
 under $\sigma^+$ circularly polarized excitation (indicated in Fig. \ref{fig:img1}b), the radiative decay
 originating from the 
 excited $5d^1$ level ends up in both ground state spin levels with equal probability. 
A repeated excitation of the selectively driven spin-flip transition eventually results in a polarized
ground state spin level. 
When applying the optical polarization procedure,  
the spin is pumped into the optically 'dark' state\cite{siyushev2014coherent}, which results in a reduced 
fluorescence signal.
 MW radiation resonant with the ground $4f^1$ state spin transition can then be used to flip the 
 optically polarized spin. This results in an increased fluorescence signal, which allows for optically
 detected 
 magnetic resonance (ODMR) measurements on individual cerium ions.

Initialization measurements, shown in Fig. 
 \ref{fig:img2}a., capture the dynamic evolution of the fluorescence signal as a function of the laser 
 pulse number. Starting with a thermal polarization, the fluorescence signal drops with increasing
 number of laser pulses because of optical pumping of the spin. 
 Initialization fidelities range between 5\% and 15\%, depending on the 
 individual Ce$^{3+}$ ion under investigation. The relatively high background of densely
 packed Ce$^{3+}$ ions in the sample can contribute to measuring a decreased fidelity.

 In our experiments, a magnetic field with a strength of 970\,Gauss was applied parallel to 
 the optical beam and the $b$-axis of the YSO crystal, for which Ce$^{3+}$ has a $g$-factor of 
 $g_{\mbox{\tiny Ce}}\approx1.4$ and a magnetic resonance frequency of
 1930.5\,MHz\cite{pidol2006epr}. MW
 structures created in proximity of SILs were then used to sweep the MW radiation frequency in order to 
 observe
  ODMR spectra of individual Ce$^{3+}$ ions. A typical spectrum is shown in Fig. \ref{fig:img2}b and
  exhibits a linewidth of 2-3\,MHz, slightly different for different cerium ions.

 Coherent Ce$^{3+}$ spin manipulation is demonstrated in a measurement of spin Rabi oscillations under
 strong MW driving, shown in Fig. \ref{fig:img2}c. 
 A typical sequence for such a Rabi measurement contains 
 spin initialization, control and readout, as shown in the inset of Fig. \ref{fig:img2}c. An exponantially 
 decaying sine function was fitted to the Rabi measurement signal, revealing a Rabi frequency
 of 5.6\,MHz and a decay time of 2\,$\mu$s.
 
 A free induction decay (FID) experiment is capable to reveal the electron spin coherence in the
 thermal noise of the characteristic nuclear spin bath of YSO, featuring $^{89}$Y and $^{29}$Si nuclear
 spins. Fig. \ref{fig:img2}d presents a typical FID of Ce$^{3+}$ and
 quantifies the inhomogenous broadening of Ce$^{3+}$ spin transition to $T_2^*=310$\,ns, 
 by fitting with a Gaussian.
 These dephasing times are
 in good agreement with ODMR linewidth measurements. The main contribution to inhomogenous broadening
 may emerge from the nuclear spin bath and can also come
 from short-pulsed optical excitation of Ce$^{3+}$, causing ionization of 
 electron traps nearby the electron spin, which induce charge fluctuations capable of stark-shifting the 
 resonance. 
   
 \begin{figure}[htb!]
\includegraphics[width=\columnwidth]{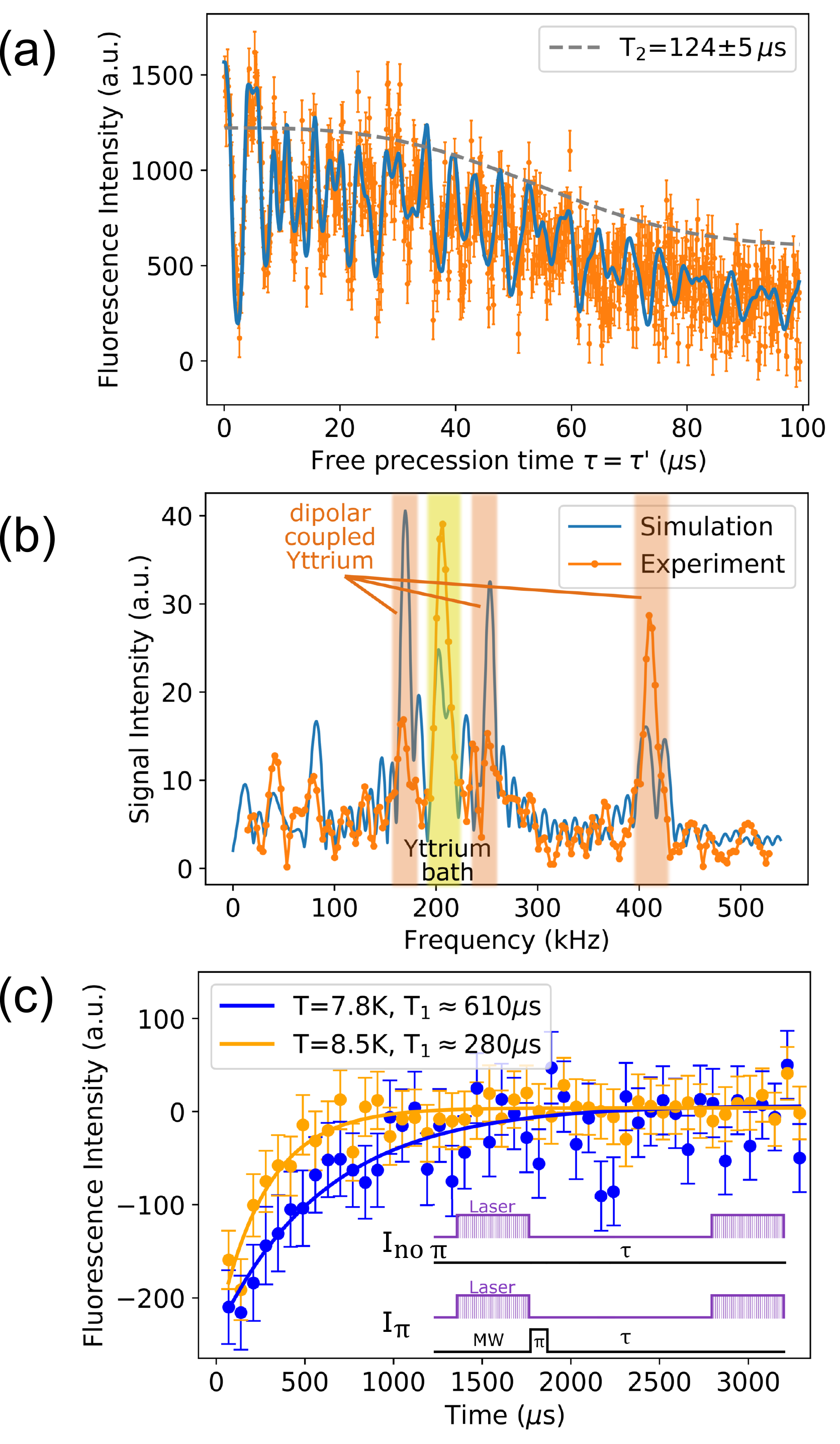}
\caption{a) Hahn echo measurement shows collapse and revivals. The decay
is fitted to $\exp[-(2\tau/T_2)^3]$ (grey dashed line), with $T_{2}\approx 124\,\mu$s, plotted offset on the 
intensity axis to envelope the signal. Simulated coherence is plotted as blue continuous line. 
b) FFT of the Hahn-Echo signal (orange) and FFT of the 
simulated Hahn-Echo signal (blue). Highlighted in yellow is the yttrium bath contribution at
$\approx 200\,$kHz. Highlighted in orange are the contributions orginiating from dipolar
coupled yttrium nuclear spins in close proximity to Ce$^{3+}$ \cite{schweiger2001principles}.  
c) Spin relaxation measurement shows $T_{1}=610\,\mu$s and $T_{1}=280\,\mu$s for sample 
temperatures of 7.8\,K and 8.5\,K. 
Signal is the difference between the two 
sequences, indicated in the inset. 
\label{fig:img3}}
\end{figure}

 The Hahn spin echo sequence decouples the spin from slow (compared to $\tau$) changes in the environment 
 and allows for more detailed spin spectroscopy of the nuclear spin environment of Ce$^{3+}$ electron spins. 
 A respresentative Hahn echo measurement on Ce$^{3+}$ is shown in Fig. \ref{fig:img3}a.
 We observe periodic revivals
 related to yttrium ions in the crystal, which have 100\% abundance of nuclear spin $I_{\mbox{\tiny Y}}=1/2$ 
 with magnetic
 moment $\mu_{Y}= -0.137 \mu_N$, with $\mu_N$ as the nuclear magneton. 
 The overall decaying signal corresponds to a decoherene time of $T_{2}=124\,\mu$s and was fitted to
 $\exp[-(2\tau/T_2)^3]$ \cite{childress2006coherent}. 
 Based on the cluster-correlation
 expansion method \cite{zhao2012decoherence}, the coherence of the electron spin in an interacting bath
 is simulated and plotted as blue line in Fig. \ref{fig:img3}a.  For the Ce$^{3+}$ electron spin
 located in a complex nuclear environment such as the YSO matrix, the simulated behaviour of its
 coherence 
  under spin echo control agrees strikingly well with our experimental results. 
  Details of the simulations are described in the supplementary material \cite{supplement}. 
  The Fast Fourier Transform (FFT) of the Hahn echo signal (see Fig. \ref{fig:img3}b)
 reveals two particular facts. Firstly, yttrium bath related signatures, found at the Lamor precession 
 frequency 
 $\omega_{\mbox{\tiny Y, Lamor}}\approx 200\,$kHz, similarly known from single NV centers in diamond
 \cite{childress2006coherent}. Additionally, we find frequency components related to weakly coupled
 yttrium nuclear spins based on magnetic dipole interaction with the electron spin. The measured 
 hyperfine coupling
 strengths are in good agreement with simulated splittings (in the range between 50\,kHz and 120\,kHz).

 We measured the lifetime of the spin state $T_1$ after optical initialization by reading out the spin
 state 
 after a variable waiting time $\tau$.
 The spin relaxation 
 measurements are shown in Fig. \ref{fig:img3}c and reveal $T_{1}=610\,\mu$s at a temperature 
 of $T=3.8\,$K, 
 measured at the heat exchanger of our cold-finger cryostat. 
 At an indicated temperature of $T=4.5\,$K, the spin lattice relaxation measurement 
 yields $T_{1}=280\,\mu$s.
 By comparing our measured $T_{1}$ values with EPR based 
 values \cite{kurkin1980epr}, we can identify the actual 
 sample temperatures to be approximatly 4\,K higher than measured on the heat exchanger, because
 of insufficient cooling power. 

 \begin{figure*}[htb]
\includegraphics[width=\textwidth]{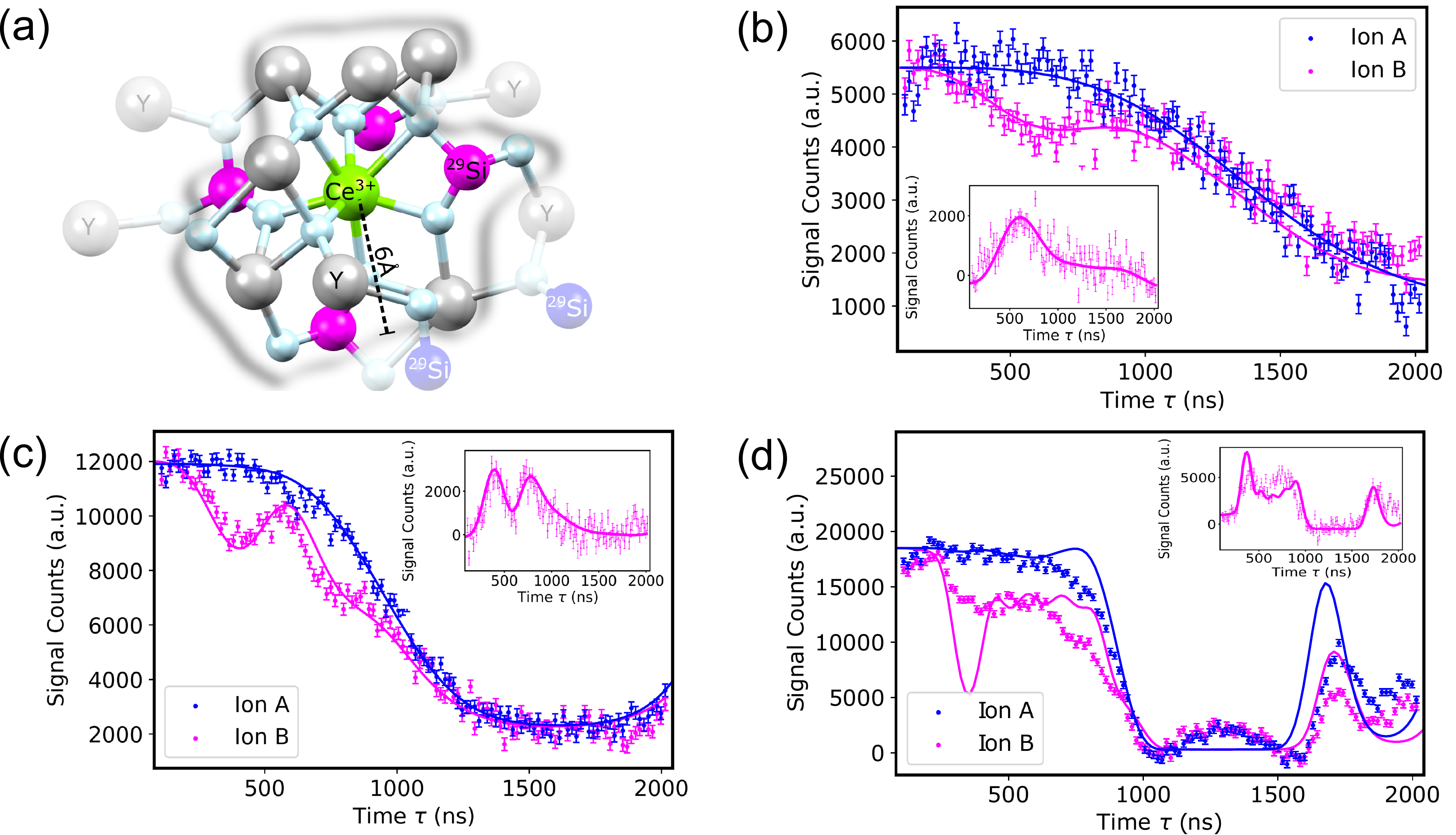}
\caption{(a) Schematic of the YSO crystal with embedded Ce$^{3+}$. Within 6\,\AA\,\,distance, 
$^{29}$Si coupling is detected (magenta), more remote $^{29}$Si cannot be distinguished (blue).
(b)-(d) CPMG-$N$ decoupling sequences with varying number $N$ of $\pi$-pulses. "Ion B" shows signal 
related 
to a $^{29}$Si nuclear spin
close to the investigated Ce$^{3+}$ ion. "Ion A" shows no $^{29}$Si signature.
Solid lines come from ab-initio simulations of the corresponding DD sequences. 
 (b) CPMG-1 (c) CPMG-2 (d) CPMG-5. Insets show differences between "ion A" and "ion B" data and simulations
 and represents only the silicon signal without overlapping yttrium modulation.
 \label{fig:img4}} 
\end{figure*}

 Using confocal microscopy, we can study spatially resolved Ce$^{3+}$ ions and their unique environment. 
 For each silicon ion in the crystal lattice, there is a 4.7\,\% natural abundance of $^{29}$Si isotope 
 with a nuclear spin $I_{\mbox{\tiny Si}}=1/2$ and $\mu_{Y}= -0.555 \mu_N$. For a Ce$^{3+}$ ion under 
 investigation, this leads to a chance of approximately 20\,\% to have a $^{29}$Si as nearest neighbor. 
 At the nearest neighbor location, the close distance ($\leq6\,$\AA) between electron spin and $^{29}$Si
 nuclear 
 spin leads to a detectable hyperfine coupling based on magnetic dipole interaction, superimposed
 on the hyperfine interaction with closeby yttrium spins and the yttrium bath 
 (schematically shown in  Fig. \ref{fig:img4}a).

Dynamic decoupling (DD) of the Ce$^{3+}$ spin from the nuclear bath allows to extract $^{29}$Si related 
signatures. Carr-Purcell-Meiboom-Gill (CPMG) control sequences were used to acquire noise spectra
for two different Ce$^{3+}$ ions, as shown in Fig. \ref{fig:img4}(b)-(d).
The center of the 
coherence dip (in Fig. \ref{fig:img4}b and inset) at $\tau_{\mbox{\tiny dip}}\approx 600$\,ns,
corresponds to a revival time of
$\tau_{\mbox{\tiny r}}=2\tau_{\mbox{\tiny dip}}$. $1/\tau_{\mbox{\tiny r}}$ matches well with the 
gyromagnetic ratio 
$\gamma_{Si}=-8.465$\,MHz/T, thus confirming the nearby $^{29}$Si nuclear spin.
Specifically, CPMG-$N$ decoupling sequences were used, with $N=1,2,5$ denoting the number of $\pi$
pulses used in the sequences. 
The bandwidth depends on the number of pulses and scales approximately with $f/(N/2)$ at frequency $f=1/2\tau$. 
 \cite{zhao2011atomic}. 
With an increasing number $N$ of $\pi$ pulses, the noise filter function 
causes the $^{29}$Si signature to be split in $N$ coherence dips. This 
can be seen in the insets of Fig. \ref{fig:img4}(b)-(d), where the $^{29}$Si signal is isolated by plotting
the difference between signals from the two Ce$^{3+}$ ions under investigation.
Solid lines in Fig. \ref{fig:img4} are simulated noise spectra of Ce$^{3+}$ ions located in
a specific nuclear spin environment.
In order to account for differences in the depth of coherence dips between experiment
and simulation, we phenomenologically add a relaxation and dephasing
mechanism to the dynamics of $^{29}{\rm Si}$ (see \cite{supplement}).
One possible reason
for a reduced depth of the coherence dip in the experiment is
external noise for proximal $^{29}{\rm Si}$. Broadening mechanisms can be related to
magnetic field fluctuations introduced by MW manipulation of Ce$^{3+}$,
the optical initialization as described for the NV center system \cite{wang2015theory}
or residual RE Kramers ion impurities in the crystal
(such as Er$^{3+}$, Gd$^{3+}$,$\ldots$), acting as a noise source on a short timescale.\\
Based on our simulations, we can localize a coupled 
$^{29}$Si nuclear spin within the nearest neighbor position in the lattice ("ion B", magenta).  
Furthermore,
the comparison spectrum without $^{29}$Si nuclear spin signatures ("ion A", blue) reveals information 
about the absence of $^{29}$Si within 6\,\AA\,\,distance from the Ce$^{3+}$ ion. The probability
to find a Ce$^{3+}$ ion with the same nuclear spin environment as "ion A" in YSO is
 $\sim 70$\%.

In conclusion, we show coherent control of an individual Ce$^{3+}$ electron spin in a YSO matrix. Using 
spin decoupling techniques, our spin spectroscopy reveals the single REI electron spin to be dipolar coupled
to 
nearby nuclear spins. Based on high density of yttrium in the host crystal, the 
future challenge is to distinguish spectrally
between individual coupled nulcear spins. 
Carefully designed DD sequences can improve detection of nuclear spin signals, which tend to be
submerged by the noisy spin bath \cite{ma2015resolving}.
$^{29}$Si nuclear spin sensing was demonstrated,
for $^{29}$Si being located within the nearest neighbor shell. 

Nuclear spins (either $^{29}{\rm Si}$
nuclear spin or yttrium nuclear spin) in proximity to the Ce$^{3+}$ electron
spin are quantum resources for quantum memory protocols.
By establishing polarization transfer techniques, for example based on
Hartmann-Hahn double resonance techniques\cite{london2013detecting}, single
$^{29}$Si nuclear spin could be initialized and potentially used as memory.
A key
concern is the coherence time of nuclear spins in this context. Decoherence
  caused by the nuclear spin bath
 will set the ultimate limitation for the
coherence time.
Due to weak dipolar interaction between nuclear spins in YSO 
compared to the Zeeman energy at applied magnetic fields,
only the pure-dephasing
interaction 
can have significant effect on the decoherence \cite{zhao2012decoherence}.
The characteristic decoherence time scale can be estimated by the nuclear spin dipolar interaction 
and gives
 $T_{2}^{*}\sim10\,{\rm ms}$ for $^{29}{\rm Si}$
 nuclear spin and $T_{2}^{*}\sim50\,{\rm ms}$ for 
yttrium nuclear spin.

Finally, this work motivates the realization of controllable multispin quantum registers
based on single REIs embedded in the YSO matrix. Access to local nodes based on environmental spins, as
demonstrated in \cite{waldherr2014quantum},
provides functionality of quantum memories, such as error correction. Furthermore, 
presented findings are applicable to 
other Kramers ions doped into YSO, such as erbium, for which 
coherent spin control and readout was demonstrated recently
\cite{raha2019Optical}. 

\vspace{5mm}
R.K. acknowledges financial support by the DFG (Grant
No. KO4999/3-1) and R.K. and J.W. acknowledge financial 
support by the FET-Flagship Project SQUARE, the EU via SMeL and 
QIA as well as the DFG via FOR
2724. N.Z. acknowledges financial support by 
NSFC (Grant No. 11534002) and NSAF (Grant No. U1530401 and Grant No. U1730449).

\vspace{2cm}
\textbf{Supplementary Information}
\normalsize

\section*{Laser Scanning Confocal Microscope Setup}

\begin{figure}[htb]
\includegraphics[width=\columnwidth]{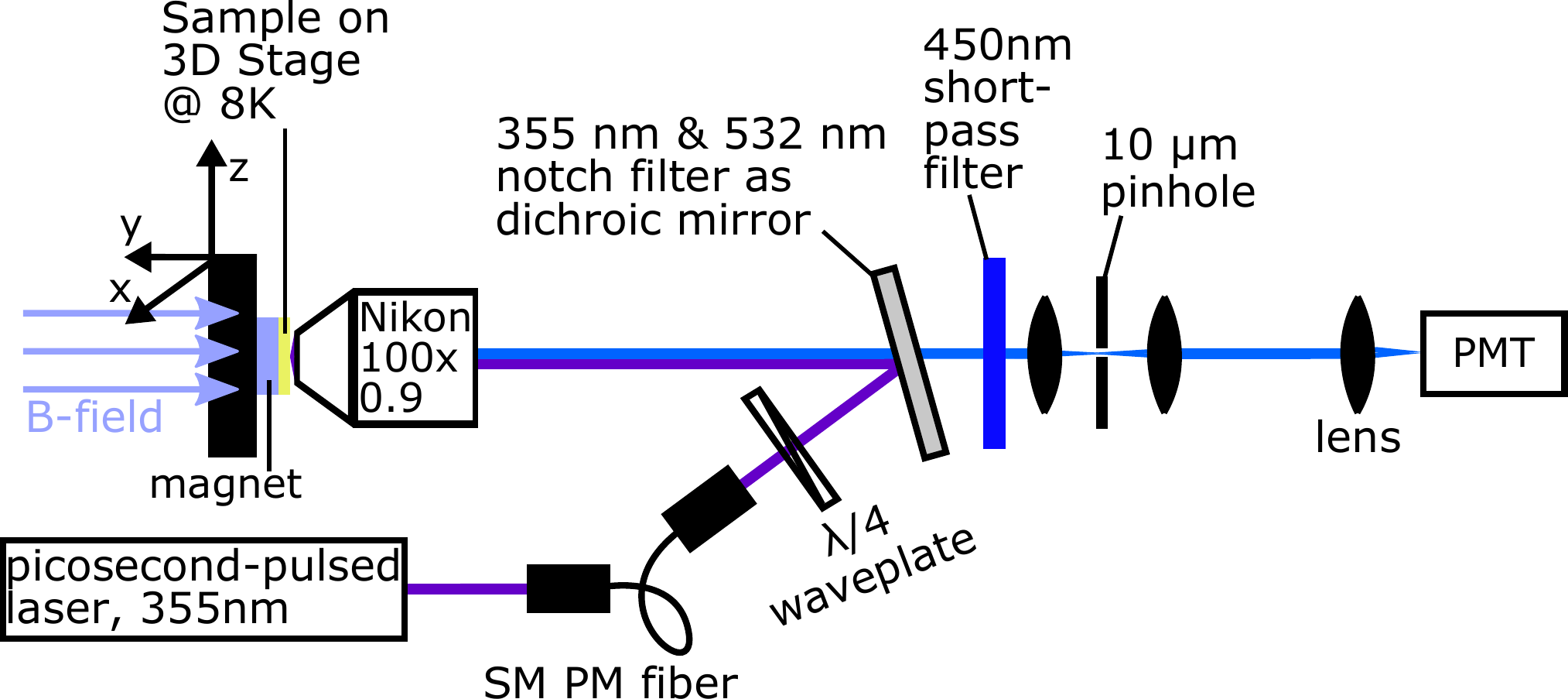}
\caption{Laser Scanning Confocal Microscope Setup. PMT: Photomultiplier Tube, SM PM fiber: single-mode 
polarization maintaining fiber. 
\label{fig:AppendixA}}
\end{figure}

The experiment was carried out in a cold-finger cryostat (CryoVac, KONTI-CRYOSTAT TYPE MICRO), where the
sample is mounted on
top of a permament magnet, which is mounted on the cold-finger 
(sketch shown in Fig. \ref{fig:AppendixA}).
The cold-finger is connected to the heat-exchanger, where also the temperature measurement is picked up. 
Due to low heat conductivity of the magnet between sample and cold-finger, the cooling power is
insufficient for thermalizing the YSO sample to the indicated temperature of $T=4$\,K, but rather 
to $T_{\mbox{\tiny sample}}=8$\,K. 
For rough positioning, the cold-finger is mounted on a 3D stage covering 10mm x 10mm x 10mm.
To obtain high resolution fluorescence images, the objective lens, mounted on a 3D piezo stage 
(NPoint, NPXY100Z25-264) in the cryostat vacuum chamber, 
can be positioned throughout 100$\mu$m x 100$\mu$m x 25$\mu$m with nanometer resolution. 

In order to obtain a picosecond-pulsed laser at around 355\,nm wavelength, we used 
a dye laser cavity, which was synchronously pumped by a 532\,nm picosecond pulsed Vanguard laser
(Spectra-Physics) with 2W output
power and 86\,MHz repetition rate. The pumped Pyridin 1 dye radiation emitting at 710\,nm was extracted 
from the laser cavity by an intra-cavity, externally triggerable Quartz AOM. With a BBO doubling
crystal, we converted the pulsed 710\,nm radiation into 355\,nm and sent it through a single-mode 
polarization maintaining (Thorlabs, PANDA PM-S350-HP) fiber for geometric mode cleaning. The linearly
polarized light after the output of the fiber was then sent through a $\lambda/4$ waveplate. It was found,
that optical polarization fidelity of Ce$^{3+}$ was not maximum at 45$^{\circ}$ tilt-angle of the 
$\lambda/4$ waveplate, corresponding to perfectly circularly polarized light, but rather at slightly
smaller tilt-angles of 40$^{\circ}$, corresponding to slightly elliptically polarized light. This
can originate from birefringence of the YSO crystal and also the 355\,nm notch filter, 
used as a dichroic mirror in the confocal microscope setup, both capable to distort the 
polarization of light, which would have to be compensated. The fluorescence of Ce$^{3+}$ was filtered
with a 450\,nm shortpass filter, in order to reject parasitic emission from impurities found in the YSO
crystal, potentially originating from other RE ion species. Additionally, flourescence 
was filtered by a 10\,$\mu$m pinhole, in order to increase the depth resolution of the microscope, before
it was collected by a Photomultiplier Tube (Hamamatsu, H10682-210).

\section*{Superresolution Microscopy}
\begin{figure}[htb]
\includegraphics[width=\columnwidth]{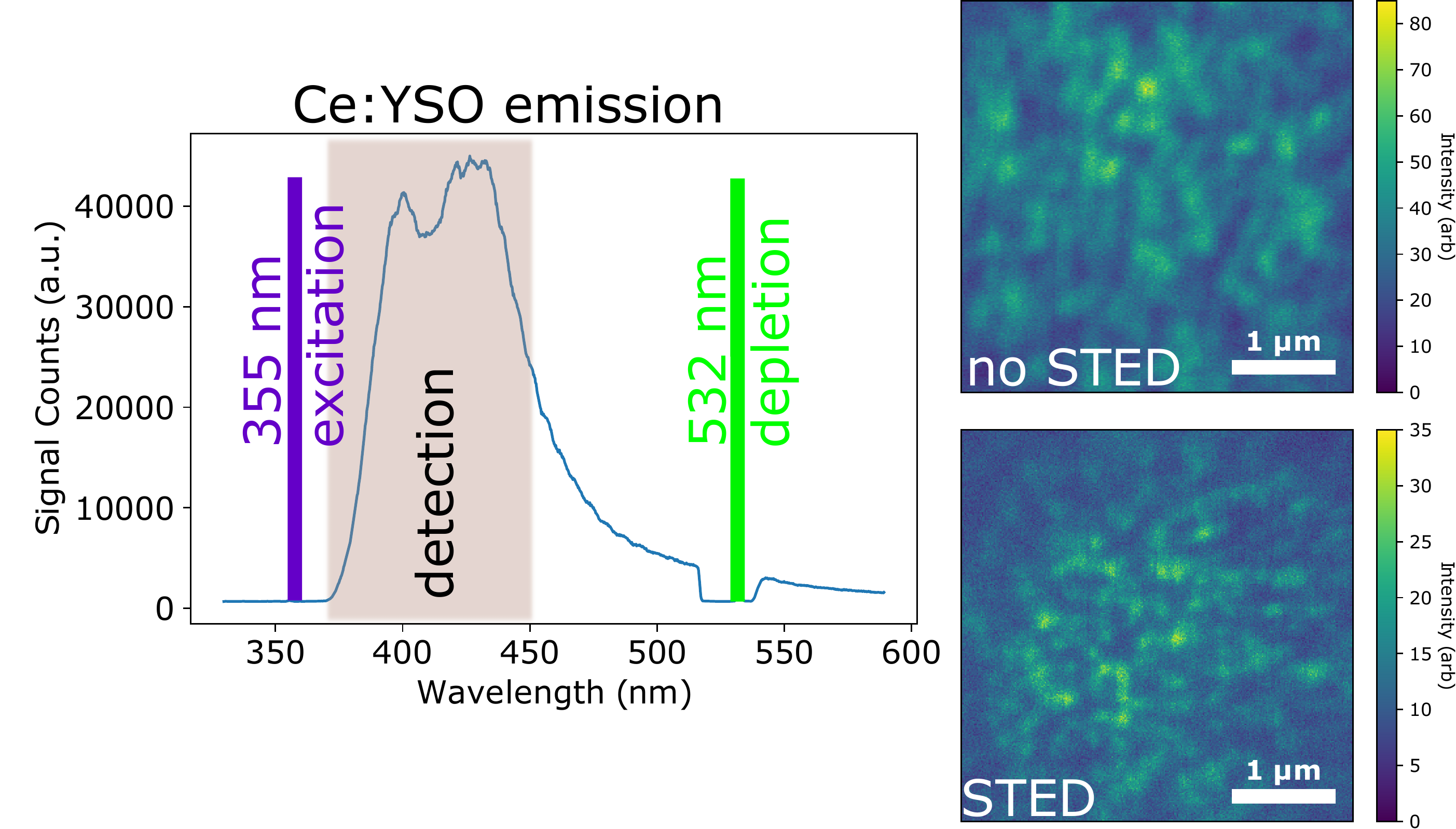}
\caption{STED microscopy of Ce$^{3+}$:YSO. Left side shows Ce$^{3+}$:YSO emission, superimposed with the 
spectral position of the picosecond pulsed 355\,nm excitation laser, the picosecond pulsed 532\,nm 
depletion 
laser and the spectral detection window for Ce$^{3+}$ fluorescence.  Right side shows two different
confocal laser scans. Top image was taken without depletion laser, bottom image was taken with depletion 
laser.
\label{fig:imgB}} 
\end{figure}

Based on similar optical properties of Ce$^{3+}$:YAG and Ce$^{3+}$:YSO, 
superresolution microscopy techniques, such as STED
microscopy, which are known to be applicable to Ce$^{3+}$:YAG\cite{kolesov2018superresolution}, were 
used to increase spatial resolution for microscopy of Ce$^{3+}$:YSO. This was achieved by superimposing
a picosecond-pulsed 532\,nm doughnut shaped depletion beam on the picosecond-pulsed 355\,nm excitation
beam. In order to obtain highest increase in optical resolution, laser powers had to be adjusted carefully
with respect to each other and the 532\,nm laser pulse arrival time had to be delayed with respect to 
the excitation pulse. The optical detection window for confocal laser scans for both, with and without 
additional STED 
beam, stayed unchanged throughout the experiments, and spanned the range between 365\,nm-450\,nm, as
indicated
in the left image of Fig. \ref{fig:imgB}. Confocal laser 
scans depicted on the right side of Fig. \ref{fig:imgB} show
on top the measurement without STED beam and on bottom with STED beam. By comparing individual
point spread functions in both scans, we can quantify a resolution enhancement of factor 2. This resolution
enhancement, though in principle capable of significantly reducing background noise for our spin
spectroscopic studies, was not available in conjunction with spin manipulation. While the 
 high power 532\,nm depletion laser was used, no magnetic resonance signals could be acquired. 
We conjecture, that high power picosecond pulses at 532\,nm prevent the optical spin initialization 
and readout, potentially through ionization of Ce$^{3+}$ and surrounding traps. 

\section*{Hahn-Echo Measurement Sequence}
\begin{figure}[htb]
\includegraphics[width=\columnwidth]{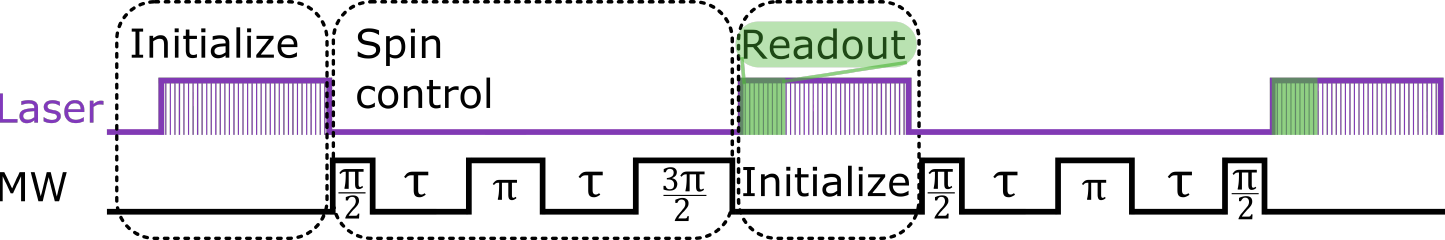}
\caption{Schematic diagram of the pulse sequence used for Hahn echo measurements,
which show the difference
between the $\nicefrac{3\pi}{2}$ MW pulse signal and the $\nicefrac{\pi}{2}$ MW pulse signal.
\label{fig:imgD}} 
\end{figure}

For Hahn-Echo measurements, the signal was acquired 
 in a balanced measurement sequence, where the final pulses in the sequence alternatingly projected 
 the spin 
 to
 $+|\nicefrac{1}{2}\rangle$ or $-|\nicefrac{1}{2}\rangle$, by applying
 either $\nicefrac{\pi}{2}$ or $\nicefrac{3\pi}{2}$ MW pulses at the end of the sequence, depicted 
 in Fig \ref{fig:imgD}b. These two 
 signals are subsequently 
 subtracted from each other, which corresponds to the coherence of the electron spin
 at a time $\tau$, defined as the average value of the transverse spin component 
 $S_t = (+|\nicefrac{1}{2}\rangle)- (-|\nicefrac{1}{2}\rangle)$. 

\section*{Simulation}

In this part, we give the simulation method of the spin decoherence
of Ce$^{3+}$ ion subjected to a YSO nuclear spin bath. The Ce$^{3+}$ ion in
the YSO crystal, substituting one yttrium, forms a defect
center. Due to the crystal field, the fine structure of the Ce$^{3+}$ ion
splits further to three doublets. The ground doublet state subspace
is our key concern. Its Hamiltonian is

\begin{equation}
H_{{\rm Ce}}=\frac{\mu_{B}}{\hbar}\boldsymbol{B}\cdot g_{{\rm eff}}\cdot\boldsymbol{S},
\end{equation}

where $\mu_{B}$ is the Bohr magneton, $\hbar$ is the Planck constant,
$\boldsymbol{B}$ is the magnetic field, $\boldsymbol{S}$ is the
lowest doublet state spin operator, and $g_{{\rm eff}}$ is the effective
g-factor of Ce$^{3+}$ \cite{pidol2006epr,wen2014spectroscopic}

\begin{equation}
g_{{\rm eff}}=\left(\begin{array}{ccc}
0.6514 & 0.2629 & 0.3004\\
0.2629 & 0.6799 & -0.0858\\
0.3004 & -0.0858 & 0.9098
\end{array}\right).
\end{equation}

The Ce$^{3+}$ ion is surrounded by a dense nuclear spin bath, which consists
of $^{89}{\rm Y}$ nuclear spins and $^{29}{\rm Si}$ nuclear
spins. These nuclear spins interact with Ce$^{3+}$ ion through magnetic
dipole-dipole interactions \cite{siyushev2014coherent}

\begin{equation}
H_{{\rm int}}=\sum_{i}\frac{\mu_{B}\mu_{0}\gamma_{i}}{4\pi r_{i}^{3}}
\boldsymbol{S}\cdot g_{{\rm eff}}\cdot\left(3\boldsymbol{n}_{i}\boldsymbol{n}_{i}-\mathbb{I}\right)
\cdot\boldsymbol{I}_{i},
\end{equation}

where $\mu_{0}$ is the magnetic constant, $\gamma_{i}$ is the gyromagnetic
ratio of the $i$-th nuclear spin, $\boldsymbol{I}_{i}$ is the $i$-th
nuclear spin operator, $r_{i}$ is the distance between the Ce$^{3+}$ defect
and the $i$-th nuclear spin, and $\boldsymbol{n}_{i}$ is the corresponding
unit vector. The nuclear spin bath itself is also governed by Hamiltonian

\begin{equation}
H_{{\rm bath}}=\sum_{i}\gamma_{i}\boldsymbol{B}\cdot\boldsymbol{I}_{i}+\sum_{i>j}
\frac{\mu_{0}\gamma_{i}\gamma_{j}\hbar}{4\pi r_{ij}^{3}}\boldsymbol{I}_{i}
\cdot\left(3\boldsymbol{n}_{ij}\boldsymbol{n}_{ij}-\mathbb{I}\right)\cdot\boldsymbol{I}_{j},
\end{equation}

where $r_{ij}$ is the distance between two nuclear spins, and $\boldsymbol{n}_{ij}$
is the corresponding unit vector.

Now, we investigate the decoherence of Ce$^{3+}$ under the dynamical decoupling
sequences CPMG-$N$, where its decoherence is mainly contributed by its
surrounding nuclear spins. In the experiment, we initialize the defect
Ce$^{3+}$ centers to a polarized state, and employ a microwave $\pi/2$
pulse to prepare the defect Ce$^{3+}$ to a coherence state. At experimental
temperatures the nuclear spins are hardly polarized so that their
initial state can be well modeled as a high-temperature mixed state
$\rho_{{\rm bath}}=\otimes_{i=1}^{N}\left(\mathbb{I}_{i}/{\rm Tr}\left[\mathbb{I}_{i}\right]\right)$,
where $\mathbb{I}_{i}$ is the identity operator for the $i$-th nuclear
spin. Driven by the Hamiltonian $H=H_{{\rm Ce}}+H_{{\rm bath}}+H_{{\rm int}}$
under CPMG-$N$, the decoherence of the defect centers can then be obtained
as \cite{zhao2012decoherence,siyushev2014coherent}

\begin{equation}
\begin{split}
 L\left(2N\tau\right)={\rm Tr}_{{\rm bath}}\left[e^{-iH_{-}\tau} \right.
\cdots e^{-iH_{+}2\tau}e^{-iH_{-}2\tau} \times\\
 e^{-iH_{+}\tau} \rho_{{\rm bath}}e^{iH_{-}\tau}e^{iH_{+}2\tau}e^{iH_{-}2\tau}
\left. \cdots e^{iH_{+}\tau}\right],
\end{split}
\end{equation}

where $\tau$ is the pulse interval of the CPMG-$N$ sequences, and $H_{{\rm \pm}}$,
defined by

\begin{align*}
H_{\pm}&=\langle\pm\vert H_{{\rm int}}+H_{{\rm bath}}\vert\pm\rangle\\
&=H_{{\rm bath}}
\pm\frac{1}{2}\sum_{i}\frac{\mu_{B}\mu_{0}\gamma_{i}}{4\pi r_{i}^{3}}\boldsymbol{n}_{B}
\cdot\left(3\boldsymbol{n}_{i}\boldsymbol{n}_{i}-\mathbb{I}\right)\cdot\boldsymbol{I}_{i},
\end{align*}

is the conditional Hamiltonian of $H_{{\rm int}}+H_{{\rm bath}}$
projected to the eigenstate of $H_{{\rm Ce}}$ $\vert\pm\rangle$
\cite{zhao2012decoherence,siyushev2014coherent}. With the help of Cluster-Correlation Expansion
(CCE) method \cite{zhao2012decoherence,yang2009quantum}, the decoherence of the Ce$^{3+}$ is numerically
calculated as

\begin{equation}
L\left(2N\tau\right)=\prod_{\left\{ C\right\} }\tilde{L}_{C}\left(2N\tau\right),
\end{equation}

where $\tilde{L}_{C}\left(2N\tau\right)$ is the decoherence of Ce$^{3+}$
induced by the nuclear spin cluster $C$. When the concerned time
scale is far less than microseconds (corresponding to interaction
strength for single nuclear spins), the decoherence of the Ce$^{3+}$ can
be well described by \textit{a non-interacting nuclear spin bath.}
In our system, we can use such a non-interacting approximation because
the Zeeman interaction strength and the hyperfine interaction strength
between Ce$^{3+}$ and the nuclear spins is much larger than the interaction
within the bath spins.

\begin{figure}[htb]
\includegraphics[width=\columnwidth]{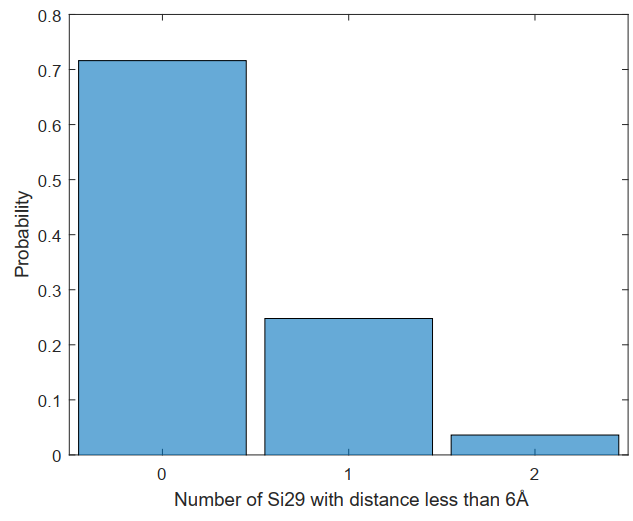}
\caption{Probability for a Ce$^{3+}$ ion in YSO crystal to have a certain
number of $^{29}{\rm Si}$ located at a distance
smaller than 6\,\AA. 
\label{fig:imgC}} 
\end{figure}

As mentioned in the main text, the cerium density in the sample can lead 
to more than one Ce$^{3+}$ ion to be probed within the focal volume.
As a consequence, the decoherence signal can in principle result from more 
than one Ce$^{3+}$ ion. The
different Ce$^{3+}$ ions in the same spot share the same yttrium nuclear
spin environment. However, different Ce$^{3+}$ ions can experience quite different
$^{29}{\rm Si}$ nuclear spin environments, based on statistcal occurance of $^{29}{\rm Si}$.
To account for this, for simulations for "ion B" shown in Fig. (4) in the main text, we assumed there
are two Ce$^{3+}$ ions in a $13\times5\times6\,{\rm nm}^{3}$ lattice but
only one Ce$^{3+}$ ion has a proximal $^{29}{\rm Si}$ with distance $3.6$
\AA . While for "ion A", we assumed there is only one Ce$^{3+}$ ion and 
no proximal $^{29}{\rm Si}$ nearby Ce$^{3+}$ in a distance smaller than 6\,\AA.
While we cannot distinguish between one or two coupled $^{29}{\rm Si}$ with distance closer than 
6\,\AA\, to
the Ce$^{3+}$ ion with the demonstrated spin spectroscopy, it is more likely to sense exactly 
one single $^{29}{\rm Si}$ nuclear spin.
Based on the 5\,\% abundance of $^{29}{\rm Si}$, we have estimated the likelyhood to detect
no $^{29}{\rm Si}$ at all, exactly 
one $^{29}{\rm Si}$ or two $^{29}{\rm Si}$, shown in Figure \ref{fig:imgC}.

The coherence difference between two different ions reflects the
difference from proximal $^{29}{\rm Si}$ nuclear spin environment.
With increasing CPMG pulse number, we expect the coherence difference
to have $N$ sharp coherence dips via a direct CCE simulation. However,
the experimentally observed coherence dips are much more shallow than
the theoretical prediction for CPMG-5. One possible reason is
external noise for proximal $^{29}{\rm Si}$, for example
magnetic field fluctuations introduced by MW manipulation to Ce$^{3+}$. To take 
noise into account, we phenomenologically add a relaxation and dephasing
mechanism to the dynamics of $^{29}{\rm Si}$ 
\cite{lindblad1976generators,breuer2002theory,wang2015theory}

\begin{equation}
\begin{split}
\dot{\rho}=\mathcal{L}_{{\rm coh}}\rho+\gamma_{2}^{{\rm Si}}
\left(\sigma_{{\rm Si}}^{z}\rho\sigma_{{\rm Si}}^{z}-\rho\right)
+\frac{\gamma_{1}^{{\rm Si}}}{2}
\left(\sigma_{{\rm Si}}^{x}\rho\sigma_{{\rm Si}}^{x}-\rho\right)\\
+\frac{\gamma_{1}^{{\rm Si}}}{2}
\left(\sigma_{{\rm Si}}^{y}\rho\sigma_{{\rm Si}}^{y}-\rho\right),
\end{split}
\end{equation}

where $\mathcal{L}_{{\rm coh}}\rho=-i\left[H,\rho\right]$, $\sigma_{{\rm Si}}^{x,y,z}$
are the Pauli matrices of the proximal $^{29}{\rm Si}$ nuclear spin,
and $\gamma_{2}^{{\rm Si}}$ and $\gamma_{1}^{{\rm Si}}$ are the
dephasing and relaxiation rates. Our simulation, shown as solid lines in the main
text in Fig. 4(b)-(d), uses $\gamma_{1}^{{\rm Si}}=\gamma_{2}^{{\rm Si}}=64\,{\rm kHz}$
(which ranges at a similar magnitude in the NV center system \cite{wang2015theory}).

\end{document}